\documentclass{article}
\usepackage{spconf,amsmath,graphicx,booktabs,url,pifont,algorithm2e,amssymb,multirow,siunitx,hyperref,cleveref}

\title{Pseudo strong labels for large scale weakly supervised audio tagging}
\name{Heinrich Dinkel,
    Zhiyong Yan,
    Yongqing Wang,
    Junbo Zhang,
      Yujun Wang}
\address{Xiaomi Corporation, Beijing, China
\\\{dinkelheinrich,yanzhiyong, wangyongqing3, zhangjunbo1,wangyujun\}@xiaomi.com
}
\begin{document}

\maketitle
\begin{abstract}
Large-scale audio tagging datasets inevitably contain imperfect labels, such as clip-wise annotated (temporally weak) tags with no exact on- and offsets, due to a high manual labeling cost.
This work proposes pseudo strong labels (PSL), a simple label augmentation framework that enhances the supervision quality for large-scale weakly supervised audio tagging. 
A machine annotator is first trained on a large weakly supervised dataset, which then provides finer supervision for a student model.
Using PSL we achieve an mAP of 35.95 balanced train subset of Audioset using a MobileNetV2 back-end, significantly outperforming approaches without PSL.
An analysis is provided which reveals that PSL mitigates missing labels.
Lastly, we show that models trained with PSL are also superior at generalizing to the Freesound datasets (FSD) than their weakly trained counterparts.
\end{abstract}
\section{Introduction}
\label{sec:intro}

Automatic audio pattern recognition is an important research topic, which enables machines to fully interact with the auditory world.
A basic task within audio pattern recognition is to distinguish between different audio event types e.g., identifying speech, hearing an explosion, which is referred to as Audio tagging (AT).
Most AT datasets are either labeled with strong supervision providing precise on- and offsets for each label or temporally weak-labeled, where tags are provided for each audio clip.
Due to the high acquisition cost of strongly supervised labels, weakly supervised labels are a common occurrence when scaling to large datasets.
The largest and most popular dataset for large-scale weakly supervised AT is Audioset, which consists of clip-level annotated 10 seconds long samples with 527 classes.
Audioset's weak labels contain many flaws, such as incompleteness (missed labels), ambiguity (``Speech'' and ``Conversation''), incorrectness, and the previously mentioned lack of access to precise timestamps\footnote{A strongly-labeled subset~\cite{Hershey2021} for 200 h (4\%) of the 5200 h exists.}.

While there exists plenty research on Audioset regarding improving performance using sophisticated models~\cite{Kong2020d,gong2021psla,gong21b_interspeech,akbari2021vatt,jaegle2021perceiver,ford19_interspeech,hong20_interspeech}, research concerned with the weak labeling problem is scarce.
Relevant work regarding label enhancement includes~\cite{kumar2020sequential}, where the authors proposed the Sequential Self-Teaching (SUSTAIN) framework, an iterative label-update framework aimed at reducing label-noise.
SUSTAIN showed a relative performance gain of up to 9\% (36.6 $\rightarrow$ 39.8) on the public Audioset evaluation dataset.
On the contrary, work in~\cite{gong2021psla} has shown that automatic label enhancement via a machine annotator leads to potential performance degradation (43.9 $\rightarrow$ 39.3) on the public evaluation set, which indicates that the presence of label-noise in Audioset is a major problem.

In our point of view, one problem worth exploring is the supervision strength, i.e., if labels provided on a shorter time resolution are superior to clip-level tags.
This work aims to ameliorate the problems of label noise and insufficient supervision by proposing a simple label augmentation framework.
Our framework uses a pre-trained machine annotator to relabel a target dataset on a smaller fixed-sized time scale.
We name our approach pseudo strong labels (PSL) since the relabeled dataset consists of soft labels on a finer time resolution.
\begin{figure}[tbp]
    \centering
    \includegraphics[width=1.03\linewidth]{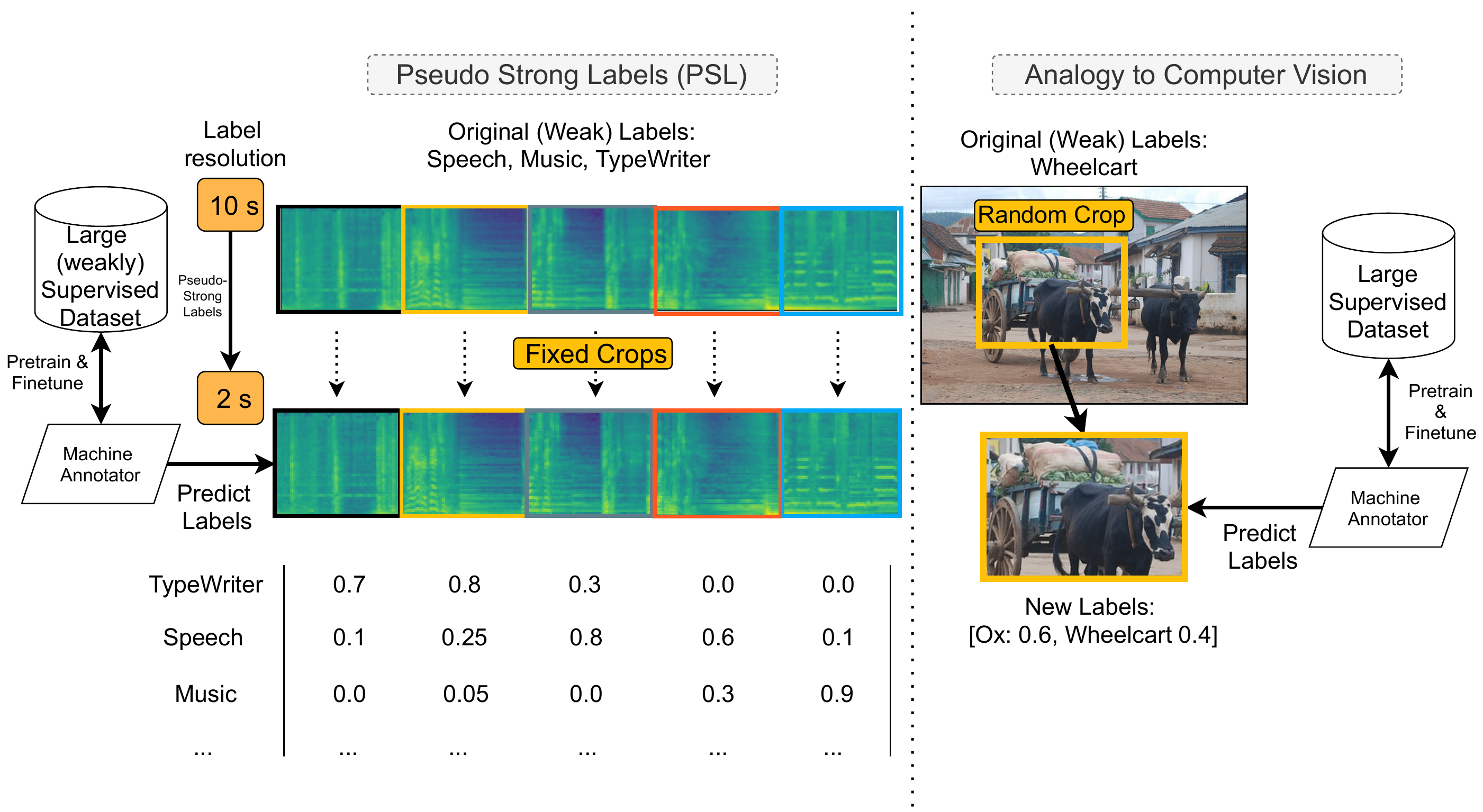}
    \caption{The proposed pseudo strong labels (PSL) framework with a comparison to ReLabel~\cite{yun2021re} in the Vision domain. A machine annotator is pre-trained on a large dataset using weak labels, which then re-estimates soft labels for fixed-sized segments.}
    \label{fig:proposed_psl}
\end{figure}
This paper is organized as follows.
\Cref{sec:method} describes our proposed approach and \Cref{sec:experiments} lays out the experimental details.
Results are presented in \Cref{sec:results} and a conclusion is provided in \Cref{sec:conclusion}.

\section{Method}
\label{sec:method}

The PSL approach pre-trains a machine annotator (MA) on a source dataset with temporally weak supervision.
Having trained the MA, soft labels are re-estimated on a target time-scale e.g., 2 s, by feeding 2 s segments to the model.
Finally, a student model is trained on this relabeled dataset.
The training objective for annotator and student is binary cross entropy (BCE) defined as:
\begin{align*}
    \mathcal{L}_{\text{BCE}}(\mathbf{x}, \mathbf{y}) &= \mathbf{y} \log{\hat{\mathbf{y}}} + (\mathbf{1}-\mathbf{y}) \log(\mathbf{1} - \hat{\mathbf{y}}), \\
    \hat{\mathbf{y}} &= \mathcal{F}(\mathbf{x}), 
\end{align*}
where $\mathbf{x}$ is an input feature of some length, $\mathbf{y} \in [0,1]^{C}$ the corresponding label, $\mathcal{F}$ the trainable neural network, and $\hat{\mathbf{y}} \in [0,1]^{C}$ the model prediction with $C=527$ classes.

We denote $\mathbf{y}_{\text{weak}}^{n}$ as a weak label provided on a time-scale $n$ (measured in seconds), with a corresponding input segment of equal length $\mathbf{x}^n$ e.g., $\mathbf{y}_{\text{weak}}^{10}$ represent the original clip-level labels in Audioset.
Note that in cases where $n<10$, we simply propagate the weak labels to match the input-segment duration  and thus optimize $\mathcal{L}_{\text{BCE}}(\mathbf{x}^{n}, \mathbf{y}_{\text{weak}}^{n})$.
Similarly, we denote $\hat{\mathbf{y}}^{n}_{\text{PSL}}$ as a soft label provided by the MA on a time-scale $n$.

MA training uses the original, weak labels $\mathbf{y}^{10}_{\text{weak}} \in \{0,1\}^C$ and training the student model uses the relabeled $\hat{\mathbf{y}}^{n}_{\text{PSL}} \in [0,1]^C$.
A description of the PSL framework is provided in \Cref{fig:proposed_psl}.
Note that PSL training discards the original labels, but we provide an ablation study of jointly using both original and predicted labels in \Cref{ssec:teacher_student}.

\section{Experiments}
\label{sec:experiments}

We explore three segment lengths for PSL: 2, 5, and 10 seconds denoted as PSL-2s, PSL-5s, and PSL-10s, respectively.
Note that PSL-$n$s denotes the model trained with labels $\hat{\mathbf{y}}^{n}_{\text{PSL}}$.
PSL-10s can be viewed as a na\"ive (weak) relabeling method, while PSL-2s and PSL-5s are the focus of this work.

\subsection{Datasets}

This work uses the largest publicly available audio tagging dataset, Audioset~\cite{gemmeke2017audio}, as the main training and evaluation corpus.
Audioset is a multi-label dataset, where a single audio clip can have up to 14 distinct labels.
Notably, the label distribution of Audioset is extremely imbalanced, where the least prevalent label ``Toothbrush'' has 67 training samples, while the most common label ``Music'' has around a million samples.
Due to partial unavailability and difficulties acquiring Audioset, we provide in-depth information about our downloaded dataset in \Cref{tab:dataset_statistics}.
We further experiment with a subset of the full training set named Aud-300h, which has been generated by sampling at most 200 audio clips for each label from the unbalanced training set and added to the balanced training set.

\begin{table}[htbp]
    \centering
    \begin{tabular}{ll||rr}
        \toprule
        Dataset & Purpose & \# Clips &  Duration (h) \\
        \midrule
        Balanced & \multirow{3}{*}{Train}  & 21,155 & 58 \\
        Aud-300h &  & 109,295 & 300  \\
        Full &  & 1,904,746 & 5244  \\
        \hline
        Eval & Evaluation & 18,229 & 50  \\
        \bottomrule
    \end{tabular}
    \caption{Dataset used in this work. The Full and Aud-300h datasets are supersets of the balanced training-set.}
    \label{tab:dataset_statistics}
\end{table}
\vspace{-4mm}
\subsection{Training setup}

Log Mel-spectrogram (LMS) features are chosen as the default front-end feature for the task.
Each 64-filter LMS is extracted from a 32 ms window with a stride of 10 ms.
If samples in a batch have an unequal duration, we apply batch-wise zero padding to the longest sample within a batch of size 32.
Adam optimization~\cite{AdamKingMa} is utilized with a starting learning rate of 1e-4.
For all experiments, every 10,000 batches ($\approx \frac{1}{6}$ epoch), we validate the model on the balanced subset with mean-average precision (mAP) as our primary metric, identical to other works~\cite{Kong2020d,gong2021psla}.
Additionally, we provide the d-prime $d'$ score, which represents our model's capability to detect the presence of an event.
The top-4 checkpoints achieving the highest mean average Precision score on the balanced dataset are weight-averaged to obtain the final model used for evaluation.
Pytorch~\cite{PaszkePytorch} was used for neural network implementation.\footnote{The source code is available online: \url{www.github.com/RicherMans/PSL}}
\vspace{-3mm}
\subsection{Model setup}
\label{ssec:model_setup}

This paper uses MobilenetV2~\cite{sandler2018mobilenetv2} as the MA and PSL student model, due to its small size (3M Parameters), quick training time, and strong performance on Audioset~\cite{Kong2020d}.
Training differences between MA and the PSL student model are provided.
\vspace{-4mm}
\paragraph*{Machine annotator}
The MA uses a pseudo-balanced sampling strategy~\cite{Kong2020d}, raw-wave (Gain, Polarityinversion, TimeShift), and SpecAug~\cite{Park2019} augmentations as well as Mixup~\cite{zhang2017mixup} as described in~\cite{Dinkel2021}.
Further, MA training uses a polynomial decay strategy over the course of the training with a duration of 30 epochs.
The MA model is trained on the full training set and achieves an mAP of 40.53.

\vspace{-4mm}
\paragraph*{Student model}
The student model is trained on the relabeled dataset provided by the MA without data augmentation.
Different from MA training, the student randomly samples the dataset, since hard labels are unavailable.
Student training is done for at most 300 epochs, with an early stop of 15 epochs.
Since evaluation data samples in Audioset are at most 10s long, we split each sample into \SI{2/5}{s}  chunks (PSL-2/5s) and average the scores obtained from the student model over an audio clip.

\vspace{-3mm}
\section{Results}
\label{sec:results}

\subsection{The effect of PSL}
\label{ssec:results_audioset}

\begin{table}[htbp]
    \centering
    \begin{tabular}{l|l|rrr}
    \toprule
        Method & Label &   mAP & $d'$ \\
        \midrule
        Baseline (Weak) & $\mathbf{y}^{10}_{\text{weak}}$ & 17.69 & 1.994\\
        \hline
        PSL-10s (Proposed)  & $\hat{\mathbf{y}}^{10}_{\text{PSL}}$ & 31.13  & 2.454 \\
        PSL-5s (Proposed) &  $\hat{\mathbf{y}}^{5}_{\text{PSL}}$ & 34.11 & 2.549 \\
        PSL-2s (Proposed) & $\hat{\mathbf{y}}^{2}_{\text{PSL}}$ & \textbf{35.48} & \textbf{2.588}\\
        \hline\hline
        CNN14~\cite{Kong2020d} & \multirow{5}{*}{$\mathbf{y}^{10}_{\text{weak}}$} & 27.80 & 1.850  \\
        EfficientNet-B0~\cite{gong2021psla} &  & 33.50 & -\\
        EfficientNet-B2~\cite{gong2021psla} &  & 34.06 & - \\
        ResNet-50~\cite{gong2021psla} &  & 31.80 & - \\
        AST~\cite{gong21b_interspeech} &  & 34.70 & -\\
        \bottomrule
    \end{tabular}
    \caption{PSL training with varying resolutions on the balanced subset of Audioset. Results shown are on the public Audioset evaluation set. The first row represents our baseline model trained with the original weak labels ($\mathbf{y}^{10}_{\text{weak}}$).}
    \label{tab:baseline_balanced}
\end{table}

We study the effects of PSL on the balanced subset and introduce our results in regards to a varying label-resolution in \Cref{tab:baseline_balanced}.
Our baseline MBv2 approach trained on the balanced subset achieves an mAP of 17.69, which can be improved to up to 35.48 when using PSL.
Notably, PSL performs favorably against other approaches in the literature using external training data~\cite{gong21b_interspeech,gong2021psla}.

If we compare PSL to the na\"ive weak relabeling method (PSL-10s), we observe an increase in terms of mAP from 31.13 to 35.48.
This improvement might stem from an increase in available data samples since PSL-2s has in fact 5 times more training samples than PSL-10s.

Recall that the MA achieved an mAP 40.53 using the full (\SI{5200}{h}) training dataset, while the PSL trained model can obtain an mAP of 35.48 (87\% of the performance) using only \SI{58}{h} (1\%) of training data.

\subsection{Comparing PSL to Teacher Student training}
\label{ssec:teacher_student}

PSL can be viewed as a special case of knowledge distillation~\cite{knowledge_distill}:
$$
\mathcal{L}(\mathbf{x},\mathbf{y}) = \alpha \mathcal{L}_{\text{BCE}}(\mathbf{x}, \hat{\mathbf{y}}) + (1- \alpha) \mathcal{L}_{\text{BCE}}(\mathbf{x}, \mathbf{y}),
$$
where $\hat{\mathbf{y}}$ is a pseudo soft label predicted from a teacher model (MA) and $\alpha = 1$.
Thus, we ask if including the original weak labels into the framework benefits performance.
We compare the previous results (\Cref{tab:baseline_balanced}) to teacher-student training with $\alpha=\{0,0.5\}$.

\begin{table}[htbp]
    \centering
    \begin{tabular}{l|l|rr}
    \toprule
    Label & $\alpha$ &  mAP & $d'$ \\
        \midrule
    \multirow{3}{*}{$\mathbf{y}^{10}_{\text{weak}}$} & \multirow{3}{*}{0} &  17.69 & 1.994 \\
     &  & {19.76} & 2.072\\
     &  & \textbf{20.21} & 2.030\\
        \hline
    $\mathbf{y}^{10}_{\text{weak}}$ + $\hat{\mathbf{y}}^{10}_{\text{PSL}}$ & \multirow{3}{*}{0.5} &  27.12 & 2.190\\
    $\mathbf{y}^{5}_{\text{weak}}$ + $\hat{\mathbf{y}}^{5}_{\text{PSL}}$ & & \textbf{28.24} & \textbf{2.273}\\
    $\mathbf{y}^{2}_{\text{weak}}$ + $\hat{\mathbf{y}}^{2}_{\text{PSL}}$ & & 28.13 & 2.203\\
    
        \bottomrule
    \end{tabular}
    \caption{Teacher-student training with varying degrees of loss contribution ($\alpha$). Experiments are trained on the balanced dataset and tested on the public evaluation dataset. Best results are highlighted in bold.}
    \label{tab:teacher_student_training}
\end{table}

The results in \Cref{tab:teacher_student_training} indicate that $\alpha=0$ enhances performance on the balanced dataset against the na\"ive \SI{10}{s} approach, which likely stems from an increase in training data sample size.
However, results with $\alpha=0.5$ indicate that the increased sample size is not necessarily a major factor, since performance between PSL-10s and PSL-2/5s only marginally improves from 27.12 to 28.24.
Overall, we observe that using standard teacher-student training is inferior to the proposed PSL, where the original labels are disregarded.
Thus, we conclude that the machine annotated labels are superior to the original weak labels.

\subsection{Analysis}
\label{ssec:analysis}

\begin{figure}
    \centering
    \includegraphics[width=\linewidth]{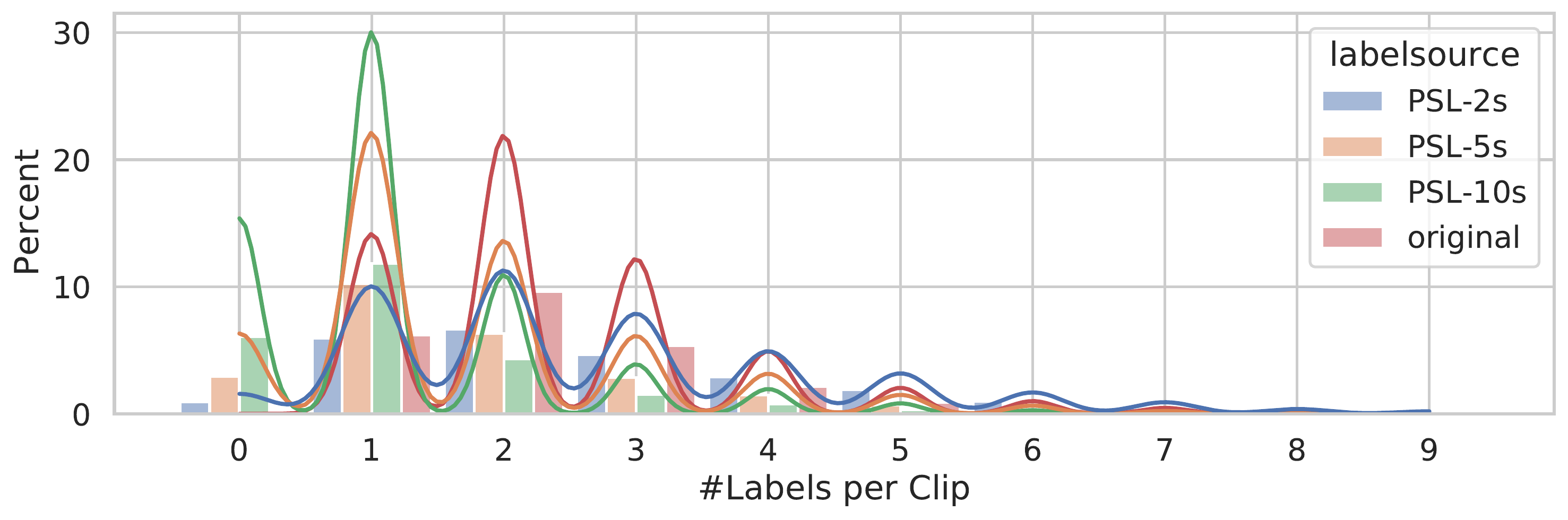}
    \caption{Number of label per clip distribution. Labels for PSL training with \SI{5}{s} and \SI{2}{s} chunks and na\"ive relabeling (\SI{10}{s}) are compared with the original (weak) labels on the balanced subset. For visualization, hard labels are obtained by thresholding the machine annotated predictions with a value of 0.5. Best viewed in color.}
    \label{fig:balanced_dist}
\end{figure}

For visualization and analysis purposes, we threshold the machine annotated labels with a value of 0.5 to obtain hard labels.
The number of labels per clip distribution is displayed in \Cref{fig:balanced_dist}.
The na\"ive PSL-10s approach mainly predicts a single label for each clip, while PSL-2s can effectively predict up to 9.
The amount of high-label per clip samples ($>4$) increases for PSL-2s compared to the original labels.
PSL-10s predicts no labels for 15\% of the training data, due to a high uncertainty, while PSL-2s fails for 3\%.
This indicates that with a stronger temporally supervision, the MA also gains a higher certainty regarding the presence of a sound event.

\begin{figure}
    \centering
    \includegraphics[width=\linewidth]{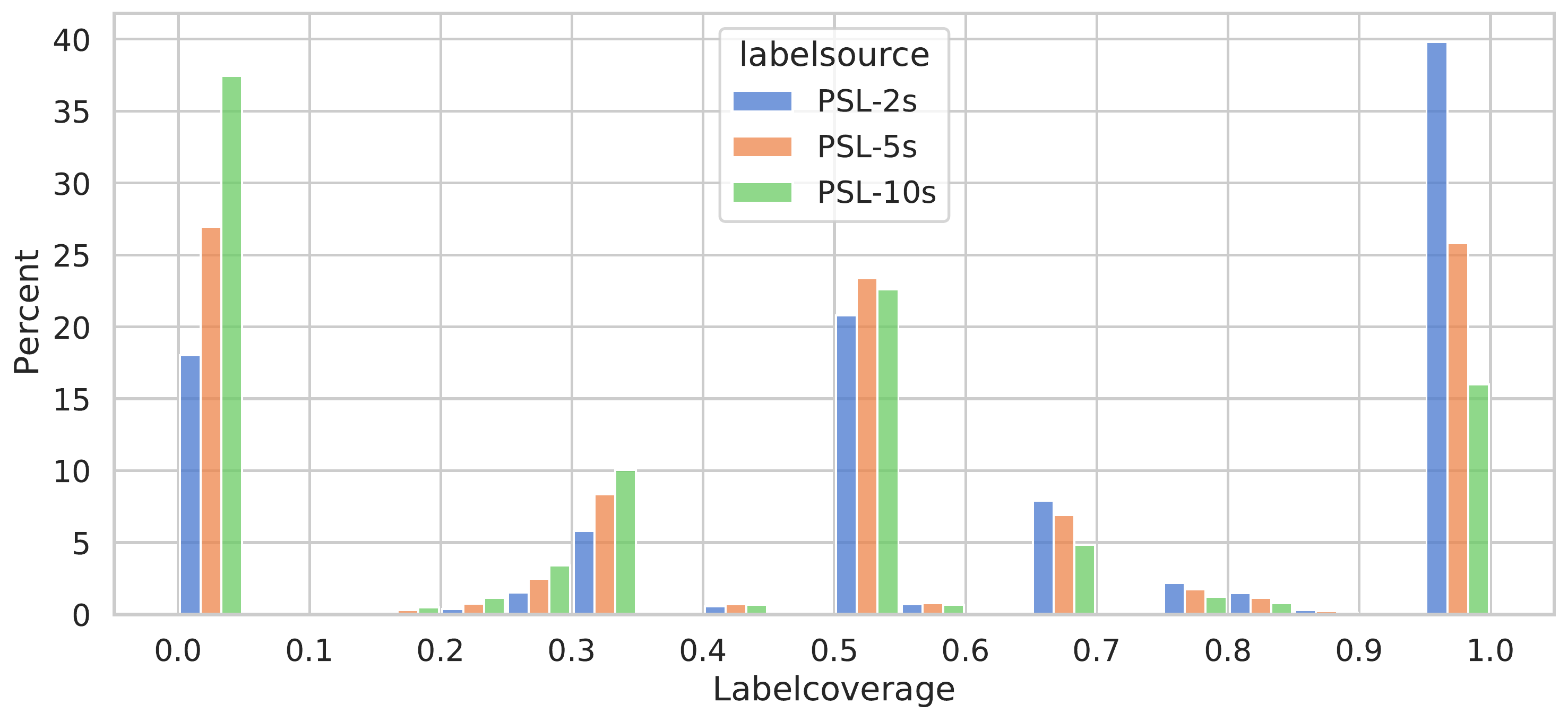}
    \caption{Labelcoverage of the machine annotated labels against the original weak labels. Best viewed in color.}
    \label{fig:balanced_coverage}
\end{figure}

Further, we analyze whether the predicted labels differ or supplement the original weak labels.
The \textit{label-coverage} between the predicted and original weak labels is shown in \Cref{fig:balanced_coverage}.
A label coverage of 1 represents that all original labels overlap with the estimated ones, while 0 represents no overlap.
The results indicate that only a fraction ($\approx$ 17 - 25 \%) of the predicted labels are entirely different from the original weak labels.
We conclude that most additional labels are supplementary to the original ones and thus partially mitigate the missing label problem.

\subsection{Training on large-scale datasets}
\label{ssec:large_scale}

We further explore whether PSL also enhances performance when used on the larger Aud-300h and full datasets.

\begin{table}[htbp]
    \centering
    \begin{tabular}{ll|rrr}
    \toprule
    Method & Training set & Label &  mAP & $d'$ \\
        \midrule
    Baseline (MA) & Full & $\mathbf{y}^{10}_{\text{Weak}}$ & 40.53 & 2.695 \\
    \hline
    PSL-10s & \multirow{3}{*}{Aud-300h} & $\hat{\mathbf{y}}^{10}_{\text{PSL}}$ & 37.97 & 2.632 \\
    PSL-5s &  & $\hat{\mathbf{y}}^{5}_{\text{PSL}}$ & 38.91 & 2.670\\
    PSL-2s &  & $\hat{\mathbf{y}}^{2}_{\text{PSL}}$ & \textbf{39.65} & \textbf{2.704}\\
    \hline
    PSL-10s & \multirow{3}{*}{Full} & $\hat{\mathbf{y}}^{10}_{\text{PSL}}$ & 38.96 & 2.710 \\
    PSL-5s &  & $\hat{\mathbf{y}}^{5}_{\text{PSL}}$ & 39.86 & {2.706}  \\
    PSL-2s &  & $\hat{\mathbf{y}}^{2}_{\text{PSL}}$ & \textbf{40.29} & \textbf{2.720} \\
        \bottomrule
        
    \end{tabular}
    \caption{Large dataset PSL training. Performance is evaluated on the public evaluation dataset. Best results are highlighted in bold.}
    \label{tab:large_scale_training}
\end{table}

The results are displayed in \Cref{tab:large_scale_training}.
Compared to previous experiments in \Cref{tab:baseline_balanced}, we notice that performance improvements are less significant.
Specifically, when comparing Aud-300h and full dataset results, we observe only marginal performance gains when using significantly more data (\SI{300}{h} vs. \SI{5200}{h}).
This behavior is likely explained by the label-imbalance in the full training set.
We provide a possible reason why performance on the full training set failed to improve, while ReLabel~\cite{yun2021re} improved performance on ImageNet.
ReLabel's MA is pre-trained on a much larger dataset (JFT-300M) compared to their target dataset (ImageNet), which is analogous to our experiments in \Cref{tab:baseline_balanced}.
Thus, we believe that PSL should lead to performance gains on the full Audioset, if a publicly available AT dataset larger than Audioset would exist.

\subsection{Transfer Learning}

Even though \Cref{ssec:large_scale} indicates that PSL cannot improve performance on Audioset via self-training, we suspect that inherent label-noise in the evaluation dataset is the main reason, as it has been previously observed in~\cite{gong2021psla}.
We, therefore, experiment on more carefully annotated weakly supervised AT datasets, such that label-noise is to some extend mitigated.
The Freesound Kaggle 2018, 2019~\cite{fonseca2017freesound,Fonseca2019audio} and FSD50k~\cite{fonseca2020fsd50k} datasets are chosen for this experiment.
Here, we use two pre-trained models, namely MA and PSL-2s trained on the full dataset (see \Cref{tab:large_scale_training}) and only train the final classifier layer, while freezing all other parameters.
Note that we segment the training samples \SI{10}{s} long for MA and \SI{2}{s} for PSL-2s, matching their respective training label resolution.
Thus, MA is trained with $\mathbf{y}^{10}_{\text{weak}}$ and PSL-2s trained with $\mathbf{y}^{2}_{\text{weak}}$.
Evaluation is done by feeding an entire audio clip of arbitrary length into each respective model, similar to \Cref{ssec:model_setup}.

\begin{table}[htbp]
    \centering
    \begin{tabular}{ll|rrr}
    \toprule
        Dataset & Metric & MA & PSL-2s & Imp.  \\
        \midrule
        FSD50k & mAP &  44.41 & \textbf{54.23} & +9.82 \\
        FSD2018 & mAP@3 & 87.31 & \textbf{89.21} & +1.90 \\
        FSD2019-Curated & \textit{lwl}wrap & 68.84 & \textbf{71.86} & +3.02 \\
        FSD2019-Noisy & \textit{lwl}wrap & 53.57 & \textbf{54.49} & +0.92 \\
        \bottomrule
    \end{tabular}
    \caption{Transfer Learning results on the FSD datasets, where only a linear classifier is trained. Both models were pretrained on the full Audioset training set.
    The absolute improvement (Imp.) is shown.}
    \label{tab:transfer_learning}
\end{table}

PSL consistently outperforms the MA baseline in regards to all downstream datasets, as it can be seen in \Cref{tab:transfer_learning}.
The largest performance gains are observed on the FSD50k and FSD2019-Curated datasets, since those contain the least amount of verified label-noise~\cite{fonseca2020fsd50k}.
We conclude that due to the temporally stronger supervision, PSL outperforms conventional weak labeled training.
% We therefore conclude that PSL is superior to weak label

\section{Conclusion}
\label{sec:conclusion}

This work proposed PSL, a simple framework that refines weakly supervised labels to provide stronger supervision.
Results on Audioset show that performance can be increased by providing cleaner supervision to a model trained for audio tagging.
PSL achieves an mAP score of 35.48 on the balanced training set outperforming weakly labeled approaches.
Our analysis reveals that PSL is capable of mitigating label noise and missing labels.
Lastly, we show that transfer learning also benefits from PSL, where we obtain a consistent performance improvement on four downstream datasets.

\vfill\pagebreak

\bibliographystyle{IEEEbib}
\small{\bibliography{refs}}

\end{document}